\documentstyle[11pt,newpasp,twoside,epsf]{article}
\begin{document}

\title{Unification Model of Seyfert Galaxies: Are All Seyfert 2 Galaxies Created Equal?}
\author{Hien D. Tran}
\affil{Department of Physics \& Astronomy, Johns Hopkins University,
Baltimore, MD 21212, USA}

\setcounter{page}{1}
\index{Tran, H. D.}

\begin{abstract}
Based on the analysis of a large spectropolarimetric survey of Seyfert 2 
galaxies (S2s) from the CfA and 12$\mu$m samples conducted mostly at 
Lick and Palomar Observatories, it is shown that
S2s with hidden BLR (HBLRs) are intrinsically more 
powerful than non-HBLR S2s. The positive detection of BLR in HBLR S2s 
appears to be due largely to the intrinsic strength of the hidden AGN 
nucleus rather than the lower level of nuclear obscuration or reduced 
dominance of circumnuclear starburst. The HBLR S2s, on average, share 
many similar large-scale characteristics with Seyfert 1 galaxies (S1s), 
as would be expected if the unified model is correct, while the non-HBLR 
S2s generally do not. These results strongly suggest that not all S2s 
are intrinsically similar in nature, and the HBLR S2s may be the only 
true counterparts to normal S1s.
\end{abstract}

\section{Introduction}

The AGN unified model proposes that Seyfert 2 (S2) galaxies are
basically the same class of object as Seyfert 1 (S1) galaxies
but viewed from a different direction. Direct evidence supporting
this picture came from spectropolarimetric observations that
showed broad, polarized permitted lines in many S2s, indicating 
that the broad-line region (BLR) characteristic of S1 is obscured 
from direct view, visible only in reflected light. Many other S2s, 
however, failed to show any signs of broad emission lines in their 
polarized flux spectra, suggesting that either the BLR could not 
exist, or other extranuclear factors (obscuration, starburst, 
geometry...) had rendered the polarization signals too weak to be 
detectable.

We carried out a large spectropolarimetric survey of S2s from the CfA 
and 12$\mu$m samples to identify those S2s with and without HBLRs. 
To illustrate their similarities and differences, we compared 
various observational properties among the HBLR, non-HBLR S2s and S1s. 

\section{Comparison between S1, HBLR S2, and non-HBLR S2}

Properties examined included indicators of AGN strength, such as
[O III] $\lambda$5007, hard X-ray (2-10 keV), radio (20 cm), and mid-infrared
(25 $\mu$m) luminosities, and various probes of obscuration, such as the 
Balmer decrement, X-ray column density, hard X-ray to [O III] flux ratio, 
and Fe K$\alpha$ equivalent width. We also compared their far-infrared 
luminosity, which is a good tracer of star forming activity. 
Our analysis showed the following:
i) By various measures, the obscuration of HBLR and non-HBLR S2s appears to be 
the same. ii) Star formation appears to be similar for all 3 Seyfert types. 
iii) Not only are HBLR S2s stronger than non-HBLR S2s, they are comparable in a lot of ways to S1s. iv) Non-HBLR S2 properties, on the other hand, generally do not match well with those of HBLR S2s or S1s. 
These results imply that: i) HBLRs are visible largely because their AGNs 
are intrinsically more powerful, not because of lower obscuration or reduced 
host starlight contamination, and ii) Non-HBLR S2s maybe too weak to represent 
real hidden S1s, and HBLR S2s may be the only true counterparts to normal S1s. 

\medskip
\noindent
For details, see Tran (2002, ApJ in press, astro-ph/0210262).

\bigskip

\begin{figure}[h]
\plottwo{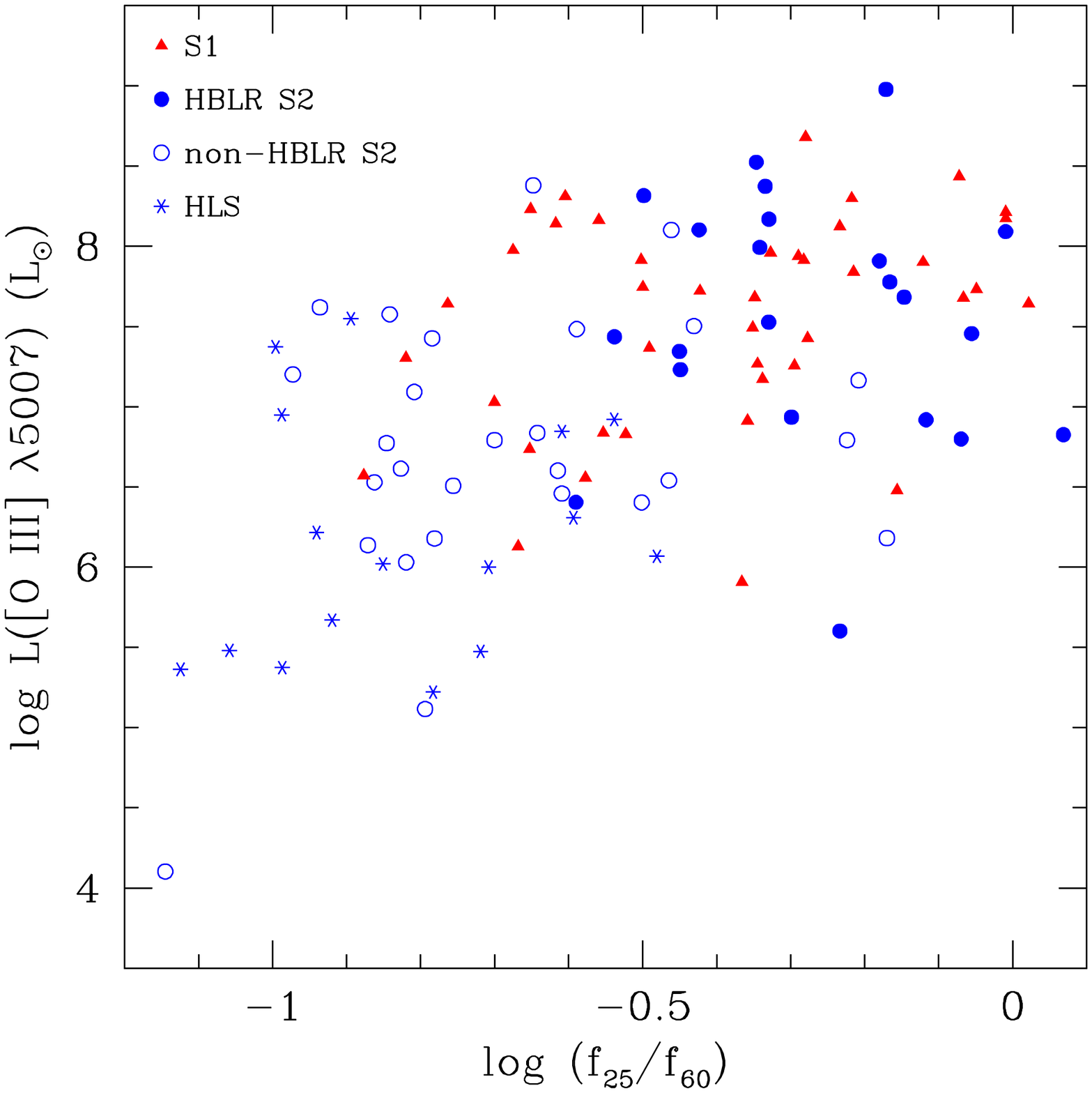}{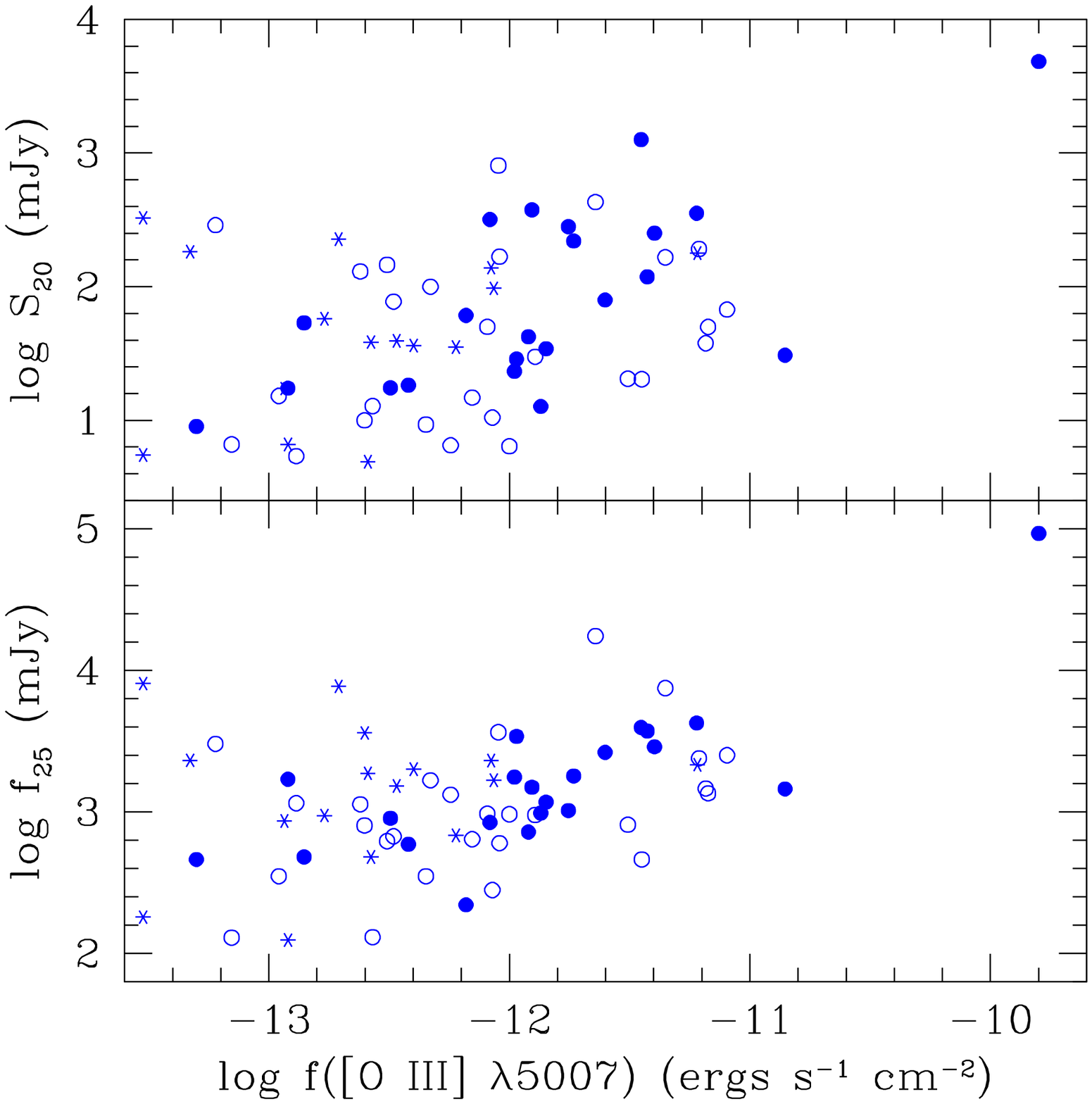}
\caption{{\it Left}: The "AGN H-R diagram": [O III]~$\lambda$5007 luminosity 
versus IR color $f_{25}/f_{60}$ for the CfA and 12$\mu$m samples. 
Good separation between HBLR and non-HBLR S2s is 
observed in this diagram. The S1s tend to lie among the HBLR S2s, while 
largely avoiding the lower left corner, which is occupied mainly by non-HBLR 
S2s and HII, LINERs, starburst (HLS) galaxies. {\it Right}: 
Extinction-corrected [O III]~$\lambda$5007, 25$\mu$m mid-IR,
and 20cm radio flux densities for the S2 and HLS
galaxies in the CfA and 12$\mu$m samples. These three properties have 
been shown to be a good indicator of the AGN strength. 
The upper-rightmost data point refers to NGC 1068. HBLR detection can reach
to very low observed flux level, below those of many non-HBLR S2s.
In addition, the flux distributions are the same for the two S2 types,
indicating that there is no preferential selection for HBLRs and the 
non-detections cannot simple all be due to the detection limit of the survey.}
\end{figure}




\end{document}